\documentclass[aps,prl,reprint,superscriptaddress]{revtex4-1}

\usepackage{graphicx}% Include figure files
\usepackage{dcolumn}% Align table columns on decimal point
\usepackage{bm}% bold math

\usepackage{xcolor}
\usepackage{ulem}
\usepackage{siunitx}
\usepackage[inkscapeformat=png]{svg}
\begin{document}

\preprint{}

\title{Rolling vesicles: From confined rotational flows to surface-enabled motion}

\author{Paula Magrinya}
\affiliation{
Department of Theoretical Condensed Matter Physics, Condensed Matter Physics Center (IFIMAC) and Instituto Nicolás Cabrera, Universidad Autonoma de Madrid, 28049, Madrid, Spain
}
\author{Pablo Palacios}
\affiliation{
Department of Theoretical Condensed Matter Physics, Condensed Matter Physics Center (IFIMAC) and Instituto Nicolás Cabrera, Universidad Autonoma de Madrid, 28049, Madrid, Spain
}
\author{Pablo Llombart}
\affiliation{
Department of Theoretical Condensed Matter Physics, Condensed Matter Physics Center (IFIMAC) and Instituto Nicolás Cabrera, Universidad Autonoma de Madrid, 28049, Madrid, Spain
}
\author{Rafael Delgado-Buscalioni}
\affiliation{
Department of Theoretical Condensed Matter Physics, Condensed Matter Physics Center (IFIMAC) and Instituto Nicolás Cabrera, Universidad Autonoma de Madrid, 28049, Madrid, Spain
}
\author{Alfredo Alexander-Katz}
\affiliation{
Department of Materials Science and Engineering, Massachusetts Institute of Technology, Cambridge, MA, 02139, USA
}
\author{Laura R. Arriaga}
\email{laura.rodriguezarriga@uam.es}
\affiliation{
Department of Theoretical Condensed Matter Physics, Condensed Matter Physics Center (IFIMAC) and Instituto Nicolás Cabrera, Universidad Autonoma de Madrid, 28049, Madrid, Spain
}
\author{Juan L. Aragones}
\email{juan.aragones@uam.es}
\affiliation{
Department of Theoretical Condensed Matter Physics, Condensed Matter Physics Center (IFIMAC) and Instituto Nicolás Cabrera, Universidad Autonoma de Madrid, 28049, Madrid, Spain
}

\date{\today}

\begin{abstract}
The interaction of surfaces in relative motion in wet environments is dominated by lubrication forces, which play a pivotal role in the dynamics of microscopic systems. Here, we develop motile vesicles that exploit lubrication forces to roll on substrates. The activity of the vesicle comes from the confined rotational flow generated by a driven rotating particle encapsulated within the vesicle by droplet-microfluidics. Lubrication forces driving vesicle rolling are controlled by membrane mechanics and its tribological properties. This provides the design principles for motile vesicles that exploit frictional forces to efficiently navigate through complex environments.
\end{abstract}

%\pacs{}% PACS, the Physics and Astronomy
  

\maketitle

Understanding the interaction of surfaces in relative motion is key for the continuous miniaturization of technological devices and biosensors. Biological systems exhibit optimized friction at the micro- and nanoscale but the underlying physical principles of their excellent performance are still poorly understood. In these systems, molecular contact is often prevented by the presence of a liquid film separating the two surfaces in relative motion, and lubrication forces that reduce friction come into play. This lubrication scenario is at the core of cell locomotion~\cite{evans2007,hoffman2011} and key to the cell internal activity~\cite{khuctrong2012}, where cytoplasmic shear flows confined by membranes may facilitate complex biological processes~\cite{goldstein2008a,vandemeent2008,Ganguly2012,Hepler2001}. Therefore, the cell membrane represents a pivotal element for both cell locomotion and force transduction of the inner cell activity into the surroundings~\cite{janmey2004}. Understanding these processes may enable the development of novel sensing mechanisms and responsive materials with enhanced biocompatibility~\cite{martina2005,fortin-ripoche2006a,sun2008}. Despite the ubiquity of membranes in natural systems and the importance of friction in natural processes, the difficulties to design controlled synthetic membrane systems capable of motion hinders our understanding ~\cite{franke2009,ewins2022}. Vesicles, aqueous drops stabilized by amphiphilic membranes, are widely used as cell membrane mimetic and may provide a versatile synthetic platform to study friction if their motion can be controlled. It is therefore essential to develop strategies to design and controllably fabricate motile vesicles as a bottom-up approach to understand lubricated friction in biological systems.

%At these low Reynolds number scenarios~\cite{purcell1977}, the motion of synthetic systems is also driven by lubrication forces. Externally-driven microparticles exploit boundary effects to translate~\cite{chamolly2020,alapan2020,fangMagneticActuationSurface2020,demirors2021,dou2021,qi2021,bozuyuk2021,bozuyuk2022,bozuyuk2022b,wu2022,vanderwee2023}. However, membranes represents a more versatile systems given their controllable mechanical properties and 

\begin{figure}[ht!]
\centering
\includegraphics[clip,scale=1,angle=0]{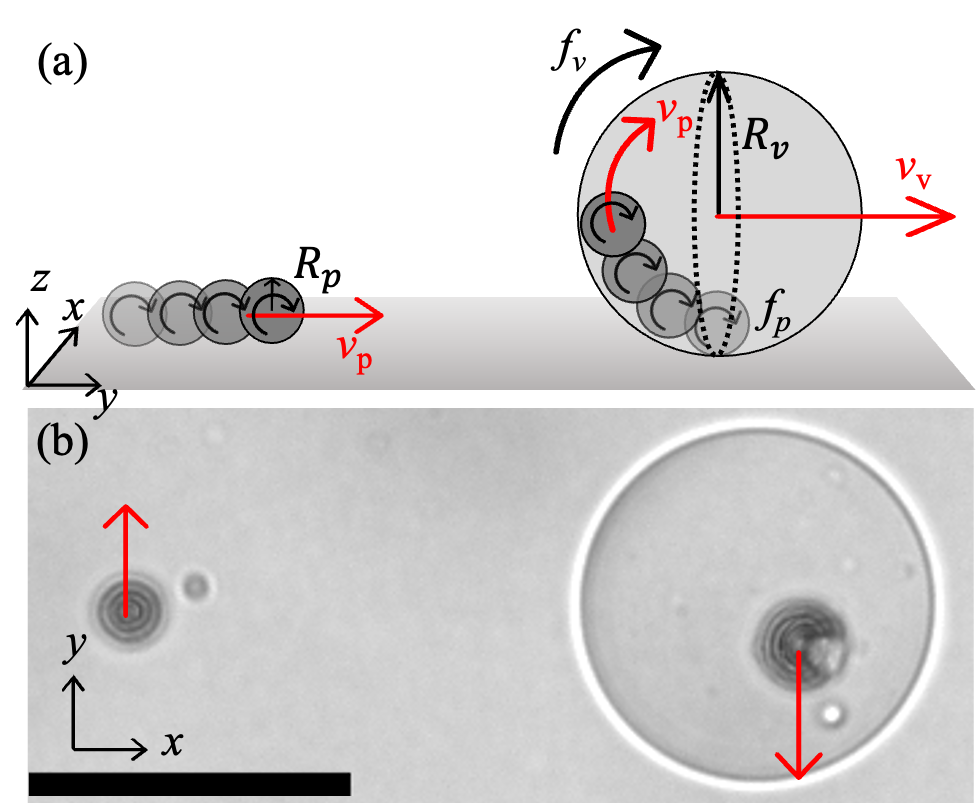}
\caption{{\bf System scheme.} (a) A particle rotating around the x-axis with frequency $f_p$ translates along the y-axis on a planar substrate, whereas it performs a circular trajectory within a vesicle, which also rotates with frequency $f_v$. (b) Bright field image showing the different direction of translation for free and confined rolling particles. Scale bar is \SI{50}{\micro\meter}.
}
\label{scheme}
\end{figure}

Here, we develop a synthetic system to study how lubrication forces enable the motion of cell-like structures by rolling on a substrate, as illustrated in Fig.~\ref{scheme}. Our experimental system comprises PEG$_{114}$-PLA$_{111}$ polymer vesicles with solid-like membranes~\cite{kim2013} and controlled radius, $R_v$, ranging from 10 to 40 $\mu$m and density $\rho_v = 1013$ kg/m$^3$. Vesicles encapsulate a single ferromagnetic particle of average radius $R_p=4 \mu$m and density $\rho_p = 1800$ kg/m$^3$ in their inner aqueous core, which is externally driven by a rotating magnetic field. To fabricate the polymer vesicles, we use water-in-oil-in-water double emulsion drops, produced with a glass capillary microfluidic device~\cite{Kim2011,shum2008}, as vesicle templates, as detailed in the Supporting Information (SI). Due to their density, the vesicles sediment in the observation chamber, whose bottom is coated, unless otherwise noted, with bovine serum albumin (BSA); this facilitates visualization of particles and vesicles by bright field microscopy. Then, we apply an external magnetic field of 10 mT, rotating around the $x$-axis, which is parallel to the substrate plane, with a frequency varying in the range 1-10 Hz, and record the particle and vesicle trajectory up to 10 min in each experiment.

Application of this rotating magnetic field to a freely suspended particle, causes particle rotation. In an unbounded fluid, the rotating particle generates a symmetric rotational flow field, with polar velocity $u_{\theta}(r)=0$ and azimuthal velocity $u_{\varphi}(r)= 2 \pi f_p R_p^3/r^2$, where $f_p$ is the rotational frequency of the particle, $R_p$ the particle radius and $r$ the distance from the particle surface. The presence of a substrate perpendicular to the $z$-axis breaks the symmetry of this rotational flow, causing particle translation with velocity $v_p$ along the $y$-axis following the direction of the shear stresses (Fig.~\ref{scheme})~\cite{goldmans1967}. Therefore, a particle rotating clockwise close to a substrate moves forward (i.e. rolling direction)~\cite{chamolly2020,alapan2020,fangMagneticActuationSurface2020,demirors2021,dou2021,qi2021,bozuyuk2021,bozuyuk2022,bozuyuk2022b,wu2022,vanderwee2023}. 
This surface-enabled rolling motion can be characterized by the dimensionless rolling parameter, $\xi_p = \frac{v_p}{2 \pi f_p R_p}$, defined as the ratio between the actual distance travelled by the particle and that expected in the case of perfect rolling. We obtain $\xi_p = 0.05$ for a free particle close to the substrate (Fig.~S3), which indicates that the coupling between particle and substrate is weak.
%and compatible with a separation distance between their surfaces of about $\delta$ $\sim$ 2 nm.

\begin{figure}[ht!]
\centering
\includegraphics[clip,scale=1,angle=0]{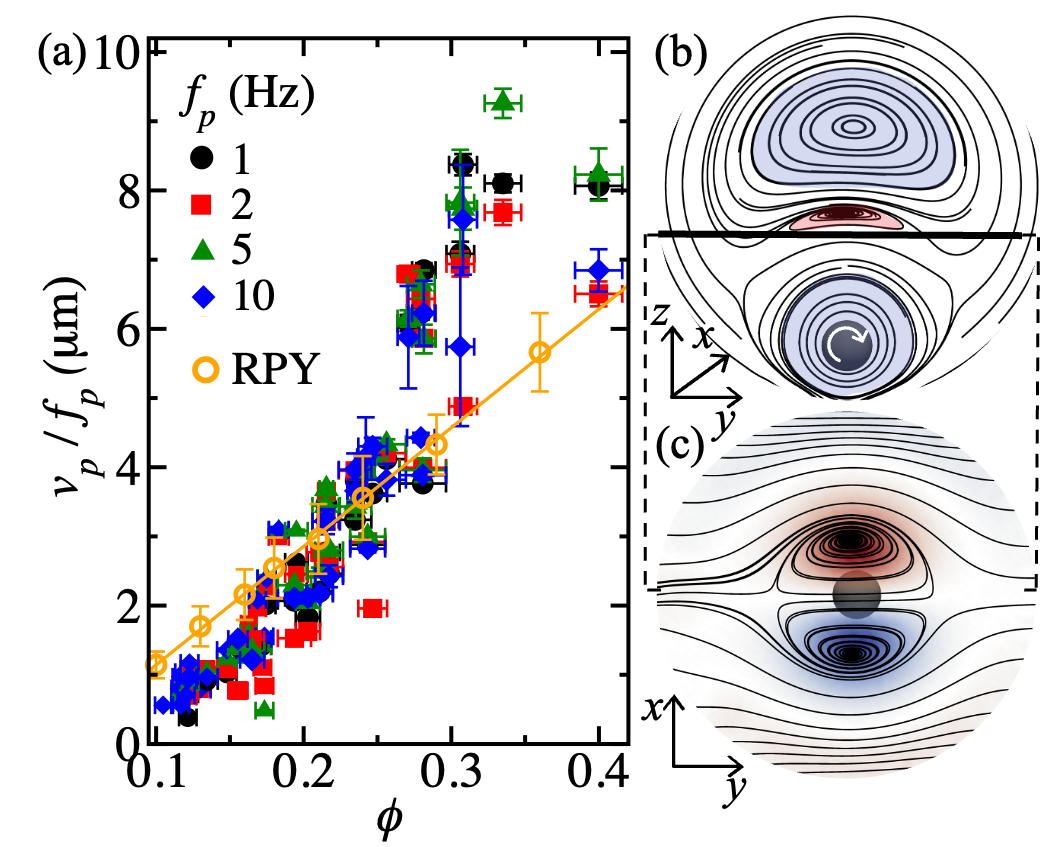}
\caption{
{\bf Dynamics of a confined rotating particle and fluid velocity fields}.(a) Distance travelled by the particle along the looping trajectory, $v_p/f_p$, as a function of the confinement, $\phi$. (b) Stream lines of the velocity field at $\phi=0.13$, obtained from RPY numerical simulations. The color map represents the vorticity around the (b) $x$ and (c) $z$ direction; positive (clockwise, red) and negative vorticity (counter-clockwise, blue).
}
\label{particle_velocity}
\end{figure}

When confined within a vesicle, the rotating particle performs a circular trajectory at the vesicle equatorial plane, defined as the $yz$-plane, as shown in the Supplementary Movie 1. For rotational frequencies above 1 Hz, these loops are sustained because the particle traverse time along the trajectory
%, $\tau_t=\frac{R_v}{v_p}$, 
is shorter than the particle sedimentation time.
%, $\tau_s=\frac{(\rho_p-\rho_s) g V_p R_v}{6 \pi \eta R_p}$ where $\rho_p$ and $\rho_s$ are the particle and solution density respectively, $g$ the gravitational acceleration, $V_p$ the particle volume and $\eta$ the fluid viscosity. 
By contrast, below this threshold frequency, we observe that the particle sediments before reaching the vesicle top (Fig.~S4). The direction of the circular trajectory described by the rotating particle within the vesicle is reversed with respect to the free particle case, as shown in Fig.~\ref{scheme}b, in agreement with a previous observation~\cite{mateos-maroto2018}. To understand this direction reversal, we measure $v_p$ as a function of the confinement, defined as the ratio between the particle and vesicle radius, $\phi = R_p/R_v$, and as a function of the actuation frequency, $f_p$. We observe that particle velocity increases with the degree of confinement and the actuation frequency, as shown in Fig.~\ref{particle_velocity}a, which indicates that the flow field established by the confinement overcomes the shear forces responsible for the rolling motion in the free particle case. In fact, the translational velocity of the particle increases linearly with the rotational frequency and thus the distance travelled by the particle in each loop, $v_{p}/f_{p}$, only depends on the degree of confinement. 

It is challenging to experimentally track the confined fluid flow in 3D. Therefore, instead, we solve the fluid flow profile by performing Stokesian Dynamics simulations of the rotating particle and the vesicle.
To this end, the particle and the vesicle are described as spherical shells of beads, connected by an elastic network of hard harmonic springs to ensure negligible deformation. Besides the excluded volume interactions between the beads forming both objects, each bead of the rotating particle is subjected to a magnetic force given by the coupling between the magnetic dipole of the particle and an external rotating magnetic field of frequency $f_p$. To determine the velocity of each bead from the forces acting on them, we use the Rotne-Praguer-Yamakawa (RPY) mobility tensor~\cite{rotne1969}. Complementarily, we carry out simulations using the Incompressible Inertial Coupling method in UAMMD\cite{uammd}, which significantly improves the description of the hydrodynamic interactions in the close contact regime (details in SI). The Stokesian hydrodynamic simulations show that the rotating particle performs a circular trajectory at the vesicle equatorial plane rolling backwards with a velocity $v_p$, shown by the orange empty circles in Fig.~\ref{particle_velocity}a, in good agreement with the experiments. The computational model also provides insight into the translating mechanism of the rotating particle. There are two opposing forces acting on a rotating particle close to a solid boundary, shear and pressure forces ~\cite{gompper2007,gompper2010,Fang2020,bozuyuk2022,caldag2022}. Shear forces are established due to the top-down symmetry breaking of the fluid shear stresses, as shown in Fig.~S9. The no-slip boundary condition imposed by the substrate makes the shear stresses at the region closer to the planar substrate higher than those at the top region and results into a net force in positive $y$-direction. On the contrary, a pressure imbalances arises from the fore-aft symmetry breaking due to the current induced by the particle rotation. The incompressible fluid dragged by the rotating particle collides with the substrate, generating a high pressure region at the right side of a clockwise rotating particle, while a low pressure region develops at the left side of the particle, as shown in Fig.~S9. This pressure difference pushes the particle in the direction opposite to rolling (i.e. sliding direction). In the case of a free particle on a planar substrate, and for the small particle surface gap of our experiments, shear forces dominate over pressure forces and the particle rolls onto the substrate in positive $y$-direction. On the contrary, above a threshold substrate curvature, pressure forces dominate and the particle translates in the sliding direction~\cite{caldag2022} (see SI for details). Experimentally, we are able to observe the exact balance between shear and pressure forces that results into vesicle rotation and translation but no net displacement of the particle, as shown in Supplementary Movie 2.

The rotating particle generates a complex vortical flow pattern inside the vesicle, as shown in Fig.~\ref{particle_velocity}b. In fixed vesicles, where membrane rotation is prevented, the circulation along any close loop over the vesicle surface is zero, and from Stokes circulation theorem, the net vorticity within the vesicle should vanish. This requires the formation of a counter-rotating vortex above the vortex created by particle rotation, which are shown in red and blue, respectively, in Fig.~\ref{particle_velocity}b. If the vesicle membrane is set free, the torque created by the particle-induced flow leads to vesicle rotation and a net circulation, which is sustained by the formation of an extra co-rotating vortex, as shown by the top blue vortex in Fig.~\ref{particle_velocity}b. When the confinement increases, the membrane and particle vortices merge (Fig.~S12). The flow structure in the $xy$-plane is strongly dependent on membrane curvature. On a planar substrate, the vorticity tube generated by a rotating particle spreads in $x$-direction. Interestingly, on a curved surface, the vorticity tube becomes tilted in the polar direction (Fig.~S13) and presents a projection in $z$-direction. The resulting vorticity pattern in the $xy$-plane is similar to a Hill spherical vortex~\cite{crowe2021}, which is composed by two vortices of opposite vorticity, as shown by the different colors in Fig.~\ref{particle_velocity}c. As the particle translates along the $yz$ vesicle equator with looping frequency $\frac{v_p}{(R_v-R_p)}$, so does the vorticity vector associated to the Hill vortex, which keeps its direction parallel to the particle position vector with respect to the vesicle center (Fig.~S14). 

\begin{figure}[h]
\centering
\includegraphics[clip,scale=1,angle=0]{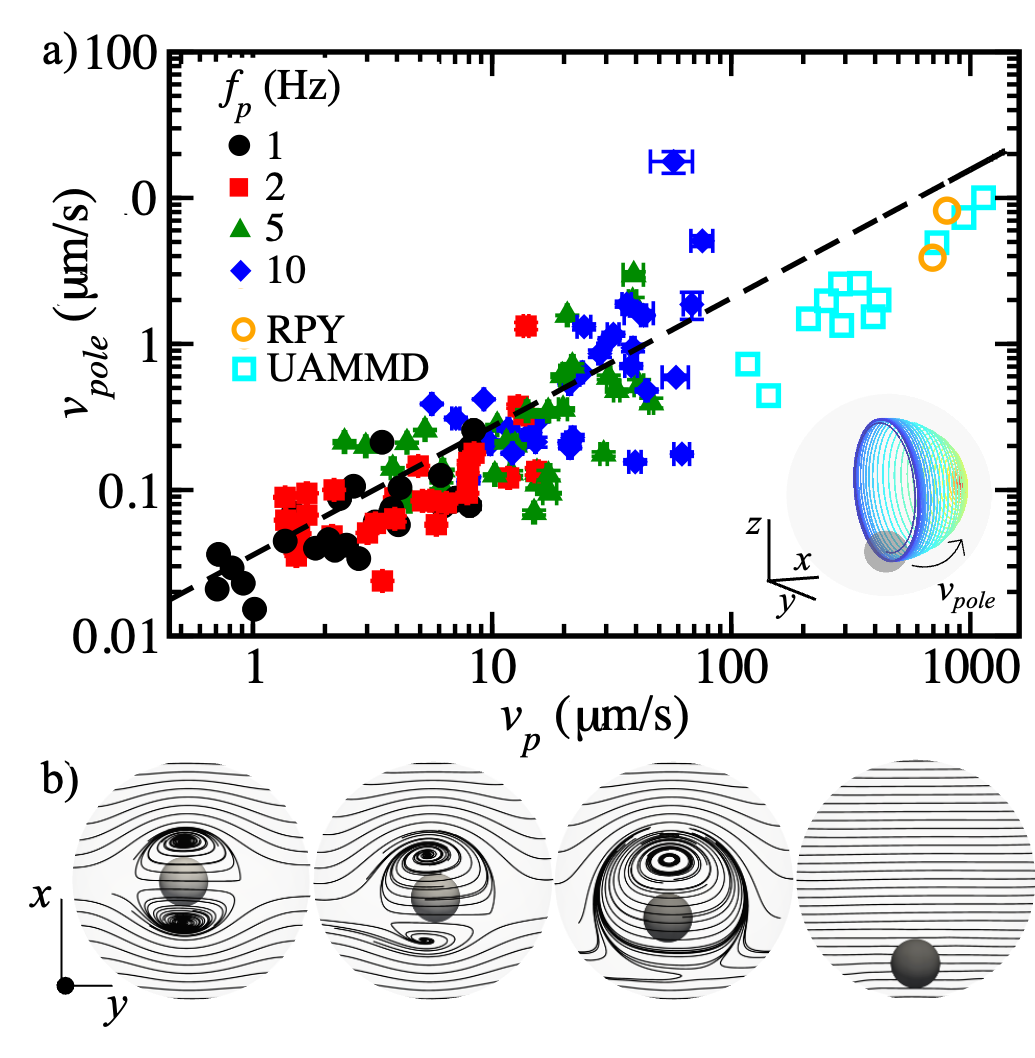}
\caption{
{\bf Particle spiraling towards the vesicle pole.} a) Drift velocity of the rotating particle towards either one of the poles as a function of its translational velocity at different rotational frequencies. The dashed line corresponds to a power law fit to the experimental results. b) Hill vortex dynamics as the particle drifts towards the vesicles pole at $\phi = 0.18$.}
\label{pole_velocity}
\end{figure}

This Hill vortex is relevant because it tends to displace the particle out from the vesicle equator. The basin of attraction for the particle equatorial trajectory is very small: any experimental or numerical disturbance will drive the particle out from the equator pushed by one of the vortices in a spiral trajectory towards either one of the stable points located at the vesicle $x$-poles, as schematically shown in the inset of Fig.~\ref{pole_velocity}a and SM1. This process breaks the symmetry of the Hill vortex in the $xy$-plane by initially enhancing the vortex opposite to the particle drifting direction, as shown in Fig.~\ref{pole_velocity}b. However, once the particle reaches the stable position at the pole, the Hill vortex structure completely disappears and the particle rotates in place until the magnetic field is switched off. The average drift velocity to the pole, $v_{pole}$, increases with $v_p$, as shown in Fig.~\ref{pole_velocity}a, but for large vesicles or very small $v_p$ the process becomes too slow and we do not observe drifting during the course of the experiment. Simulations correctly reproduce the drift velocity, yet due to limitations in computational time results are restricted to $v_{pole} > 1 \mu$m/s when using RPY and $v_{pole} > 0.1 \mu$m/s when using UAMMD. We also observe that the marginal stability of the equatorial trajectory increases when lower resolution (larger bead sizes) is used. 

\begin{figure}[h!]
\centering
\includegraphics[clip,scale=1,angle=0]{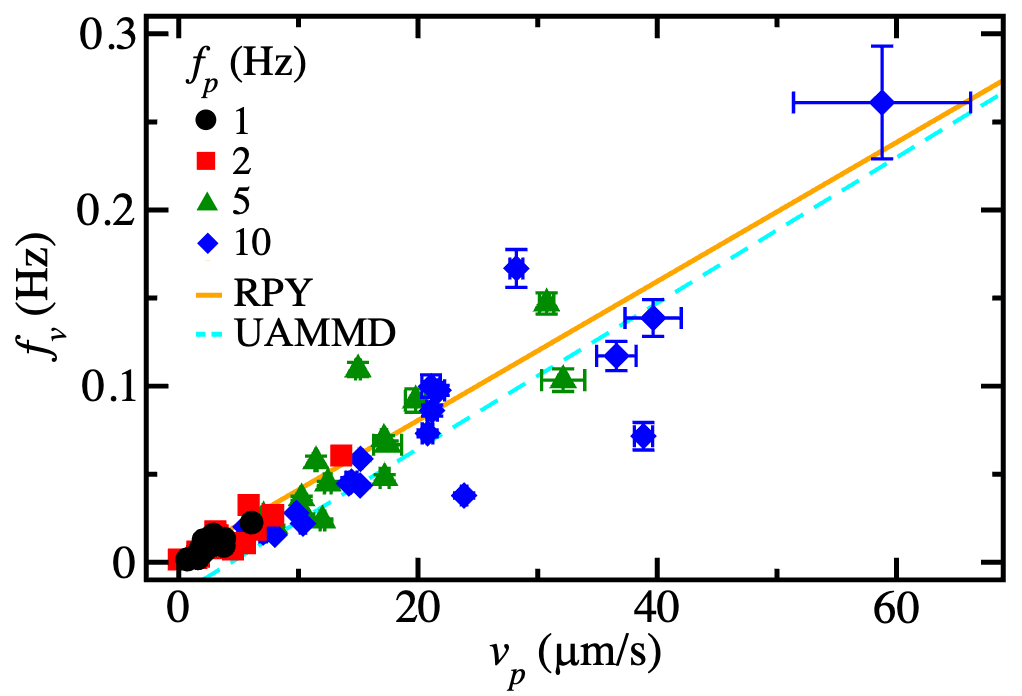}
\caption{
{\bf Vesicle rotation.} (a) Vesicle rotational frequency, $f_v$, as a function of the particle translational velocity, $v_{p}$ at increasing $f_p$. Orange and cyan lines corresponds to linear fits to the computational models results.}
\label{vesicle_rotation}
\end{figure}

Vesicle rotation is induced by the torque transmitted from the generated confined flows. We measure the rotational frequency of the vesicles, $f_v$, as a function of the translational velocity of the encapsulated particle, tracking the motion of defects on the vesicle membrane (SI). The rotational frequency of the vesicle increases linearly with particle velocity, as shown in Fig.~\ref{vesicle_rotation}, which is indicative of a linear coupling at small Reynolds number. We find that $f_v \sim$ 0.1 $f_p \phi$; this provides a simple relationship to design rotating vesicles with controlled rotational frequency. Vesicle rotation is significantly slower than particle rotation as it is essentially transferred by viscous fluid traction in a confined environment. Notably, the same dependence is obtained from the computational models. 

\begin{figure}[h!]
\centering
\includegraphics[clip,scale=1,angle=0]{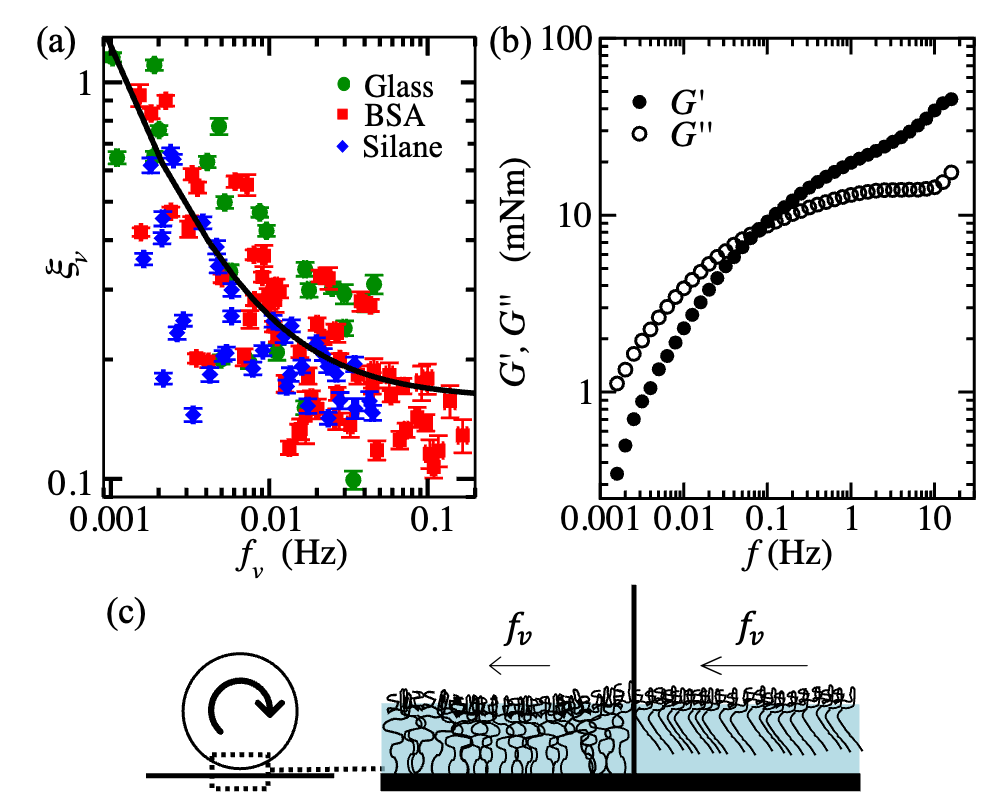}
\caption{
{\bf Vesicle translation.} a) Vesicle rolling parameter, $\xi_v$, as a function of the vesicle rotational frequency, $f_v$, on three chemically different substrates. The line corresponds to a fit to the phenomenological model. b) Frequency dependence of monolayer storage, $G'$, and loss modulus, $G''$, at a PEG-PLA surface density of 8 $\mu$g/cm$^2$ and strain 0.5\%. c) Schematic representation of the shear alignment of the tethered PEG chains as a function of vesicle frequency, $f_v$ and the appearance of a slip plane.}
\label{vesicle_translation}
\end{figure}

Rotation and translation of the vesicle are coupled through friction forces with the substrate. This coupling is given by the vesicle rolling parameter, $\xi_v=v_v/(2\pi f_v R_v)$, where $v_v$ is the vesicle translational velocity. We observe that slowly rotating vesicles translate more efficiently than those rotating at higher frequencies, as shown in Fig.~\ref{vesicle_translation}a. To elucidate if the observed transition from perfect rolling to mostly slipping is controlled by specific molecular interactions between the polymer membrane of the vesicle and the BSA-coated substrate, we repeat the experiments on uncoated and silane-coated substrates. Our experiments confirm that the observed transition is independent on the chemical nature of the substrate coating, and thus likely related to the tribological and rheological properties of the vesicle membrane. The translation of the vesicle in the $y$-direction is determined by the force balance between the rotational friction force, $F_r > 0$, controlled by the membrane-substrate interaction, and the translational friction, $F_t < 0$, given by its hydrodynamic resistance. The behavior of $\xi_v$ suggest that $F_r$ can be split into contact and lubricated friction components. The contact friction regime is governed by the microscopic contact forces between the membrane and the substrate, which are proportional to the load, $\mu N$, where $\mu$ is the static friction coefficient and $N$ is the normal force. In the lubricated friction regime, the viscous friction force is proportional to the slipping friction coefficient, $\mu_s$, and to the vesicle slip velocity with respect to the substrate, $v_{slip} = 2\pi f_v R_v - v_v$. Thus, $F_r = \mu N + \mu_s v_{slip}$. A vesicle translating at a distance $\delta$ from a substrate experiences a viscous drag $F_t=\mu_t v_v = 6 \pi \eta R_v \tilde{\mu_t} v_v$, where $\eta$ is the fluid viscosity and $\left| \tilde{\mu_t}\right|=\left|(8/15) \ln(\delta/R_v) - 0.9588)\right| \sim$ [5-8]~\cite{goldmans1967} at the experimental separation gap $\delta/R_v \sim $ [10$^{-4}$-10$^{-6}$]. The force balance yields $\xi_v = f_v^{cr}/f_v +B$, where $f_v^{cr}$ is the crossover frequency for the transition to partial slippage, $f_v^{cr}=[(\rho_v-\rho) \mu R_v g]/(9 \pi \eta \tilde{\mu})$ being $\tilde{\mu}=(\mu_s+\mu_t)/(6 \pi \eta R_v)$ and $B=\mu_s/(\mu_t+\mu_s)$. This phenomenological model captures the perfect rolling behavior, $\xi_v = 1$ for $f_v<f_v^{cr}$ for which adhesion forces are able to avoid slipping, predicting a contact friction coefficient $\mu \sim $ [0.12-0.05] for the PEG-PLA vesicle membrane independently of the substrate. For frequencies above $f_v^{cr}$=0.001 Hz, $\xi_v$ decays, reaching the lubricated friction limit at $f_v = f_v^{cr}/B =0.06$ Hz. Interestingly, at this frequency a PEG-PLA interfacial monolayer exhibits a crossover from a liquid-like ($G' < G''$) to a solid-like ($G' > G''$) behavior, as shown in Fig.~\ref{vesicle_translation}b. Below the crossover frequency, the tethered PEG chains have time to relax and thus, they interact with the substrate, resulting into a stronger friction force. Above the crossover frequency, the PEG chains become shear aligned, which may open some hydrated sliding plane where hydrodynamic lubrication prevails, resulting in a small rolling parameter of the vesicle. The model predicts a slipping friction coefficient $\mu_s \sim $ 0.17 $\mu_t$, which is not far from the purely hydrodynamic lubrication limit $\mu_s/\mu_t \rightarrow 1/3$ when $\delta \rightarrow 0$. 

The model system here developed, comprising a magnetic particle encapsulated within a polymer membrane, provides a novel tribological setup to study surface-enabled motion of cell mimicking structures in complex environments. The activity of the vesicle comes from the confined rotational flow generated by an externally-driven rotating particle. Despite the complexity of the confined flow and its transient regime, we demonstrate that vesicle motion is highly controllable. Independently on the position of the particle within the vesicle, at all times the vesicle experiences a solid-body rotation with a frequency exclusively controlled by the vesicle size and the particle rotational frequency. Rotation of the vesicle on the planar substrate then results in vesicle translation due to friction forces between the membrane and substrate. At low vesicle rotational frequencies, we access a contact friction regime, which yields perfect vesicle rolling. At higher vesicle rotation frequencies, we observe the transition from perfect rolling to mostly slipping, due to shear alignment of membrane tethered polymer chains, which reduces the contact friction between the membrane and the substrate allowing for a lubricated viscous layer. Therefore, this setup opens a relevant avenue to study the role of membranes in force transduction of internal flows and tribology in more complex membrane systems, including fluid-like lipid membranes or even membranes decorated with specific receptors to bind ligands on the substrate.

\begin{acknowledgments}
The authors acknowledge financial support for grants PID2022-143010NB-I00 and CEX2018-000805-M supported by MCIN/AEI/10.13039/501100011033/ and by "ERDF A way of making Europe". PM acknowledges grant PRE2019-091190 funded by MCIN/AEI/10.13039/501100011033/ and by "ESF Investing in your future". LRA and JLA acknowledge grants RYC2018-025575-I, RYC2019-028189-I, CNS2023-145460 and CNS2023-145447 funded by MCIN/AEI/10.13039/501100011033/ and by "European Union NextGenerationEU/PRTR". RD-B and PP acknowledge grants PID2020-117080RB-C51 supported by MCIN/AEI/10.13039/501100011033/ and by "ERDF A way of making Europe" and PDC2021-121441-C21 funded by MCIN/AEI/10.13039/501100011033/ and by "European Union NextGenerationEU/PRTR".
\end{acknowledgments}

\end{document}